\RequirePackage{fix-cm}
\documentclass[final,5p,twocolumn,authoryear]{elsarticle}
\usepackage{graphicx}
\usepackage{amsmath,amssymb,amsfonts}
\usepackage{booktabs,tabularx,array,multirow}
\usepackage{enumitem,stfloats}
\usepackage[nospread]{flushend}
\usepackage[font=small,skip=5pt]{caption}
\usepackage[table]{xcolor}
\usepackage[skins,breakable]{tcolorbox}
\usepackage[hyphens]{url}
\usepackage[unicode,colorlinks=true,linkcolor=black,citecolor=black,urlcolor=blue]{hyperref}
\definecolor{ObservationBg}{HTML}{F7F8F9}
\definecolor{ObservationRule}{HTML}{BBC2C9}
\definecolor{TableHeaderBg}{HTML}{DCEAF6}
\definecolor{TableBestBg}{HTML}{EAF3FC}
\definecolor{TableRuleColor}{HTML}{6D89A2}
\newcommand{\papertablestyle}[1]{%
  \fontsize{8}{10}\selectfont
  \setlength{\tabcolsep}{#1}%
  \renewcommand{\arraystretch}{1.18}%
  \arrayrulecolor{TableRuleColor}%
  \setlength{\heavyrulewidth}{0.6pt}%
  \setlength{\lightrulewidth}{0.35pt}%
}
\newcommand{\metrichead}[1]{\textbf{#1}}
\newtcolorbox{observationbox}[1][]{enhanced,colback=ObservationBg,
  colframe=ObservationRule,boxrule=0.45pt,arc=1pt,boxsep=0pt,
  left=5pt,right=5pt,top=4pt,bottom=4pt,before skip=5pt,after skip=6pt,#1}
\newcolumntype{Y}{>{\raggedright\arraybackslash}X}
\newcolumntype{Z}{>{\centering\arraybackslash}X}
\newcolumntype{R}{>{\raggedleft\arraybackslash}X}
\newcolumntype{L}[1]{>{\raggedright\arraybackslash}p{#1}}

\biboptions{authoryear,round,sort}
\apptocmd{\thebibliography}{\raggedright\interlinepenalty=10000\relax}{}{}
\journal{Computers \& Security}

\begin{document}
\begin{frontmatter}
\title{JANUS: Permission-Guided Intrusion Detection in Kubernetes via Cross-Modal Association}
\author[aff1]{Yongqiang Li}
% \ead{li_yq2002@tju.edu.cn}
\author[aff1]{Tao Luo}
% \ead{luo_tao @ tju. edu.cn}
\author[aff1]{Guohua Xin\corref{cor1}}
\ead{xinguohua@tju.edu.cn}
\author[aff2]{Cong Wang}
% \ead{wangcongjcdd@tust.edu.cn}
\author[aff1]{Xinke Zhang}
% \ead{zhangxinke_526@tju.edu.cn}
\author[aff3]{Jiongchi Yu}
% \ead{jiongchiyu@acm.org}
\author[aff1]{Guangquan Xu}
% \ead{Losin@tju.edu.cn}
\cortext[cor1]{Corresponding author.}
\affiliation[aff1]{organization={Tianjin University},
  city={Tianjin},country={China}}
\affiliation[aff2]{organization={Tianjin University of Science \& Technology},
  city={Tianjin},country={China}}
\affiliation[aff3]{organization={Nanyang Technological University},
  country={Singapore}}

\begin{abstract}

Kubernetes has become a widely adopted platform for deploying and orchestrating cloud-native applications that support critical services. Workload permissions govern access to Kubernetes resources, but their abuse can expose sensitive information and enable privilege escalation. Static analysis of role-based access control (RBAC) configurations can identify workloads' effective permissions and potential privilege-escalation paths, but cannot establish whether these permissions are exercised at runtime. In contrast, runtime intrusion detection based on system calls or provenance graphs can identify anomalous behavior, but lacks the authorization context needed to interpret such behavior in terms of Kubernetes permissions.

To address these limitations, we propose JANUS, a permission-guided intrusion detection method for Kubernetes that integrates static authorization semantics with runtime behavioral evidence. JANUS first identifies high-risk effective permissions and extracts their authorization subgraphs from Kubernetes configurations, then constructs permission--event training pairs through controlled execution and offline Audit verification. Using these pairs, a cross-modal model learns associations between permission semantics and runtime behavior. The resulting historical permission evidence is incorporated into provenance-graph representations through gated residual fusion, enabling permission-conditioned node-level anomaly detection.

Experimental results show that JANUS achieves an F1 score of 0.3697 and a Matthews correlation coefficient (MCC) of 0.3247 in node-level detection, the latter accounting for all four outcomes of binary classification and complementing F1 under class imbalance. Compared with detection without permission conditioning, JANUS reduces false-positive nodes by approximately 14.3\% while maintaining comparable recall.

\end{abstract}


\end{frontmatter}

\section{Introduction}

Kubernetes has become a widely adopted platform for deploying and managing cloud-native applications, including critical services~\citep{kubernetes_overview}. Its role-based access control (RBAC) mechanism governs access to Kubernetes API resources through permissions granted to workload identities~\citep{kubernetes_rbac}. While these permissions support legitimate operations, their abuse can expose sensitive information and enable privilege escalation~\citep{excessive_permissions,kubesec}.

After compromising a workload through an application vulnerability, an attacker may obtain its ServiceAccount token and use the workload's existing permissions to issue requests to the Kubernetes API server. Depending on the compromised identity's effective permissions, the attacker may read sensitive Secrets, execute commands in other Pods, or create privileged workloads, subject to applicable admission policies. If the compromised identity also satisfies the authorization requirements for creating or modifying RBAC roles or bindings, the attacker may further escalate privileges~\citep{excessive_permissions,kubernetes_rbac_good_practices}. Because these operations may be authorized despite serving malicious purposes, RBAC decisions alone cannot distinguish legitimate permission use from abuse.

Static security analysis examines these RBAC authorization relationships to identify excessive permissions, high-risk grants, and potential privilege-escalation paths~\citep{epscan,kubiscan}. However, these analyses characterize a subject's authorized capabilities rather than its actual use of permissions. The same RBAC configuration may remain unchanged regardless of whether a permission is unused, exercised legitimately, or abused. For example, neither reading Secrets nor invoking \texttt{pods/exec} directly alters RBAC authorization relationships, so configuration snapshots alone cannot establish whether these operations have occurred.

In contrast, runtime intrusion detection methods analyze system-call traces or provenance graphs to identify anomalous process, file, and network activity~\citep{threatrace,magic,kairos,flash,orthrus}. Although these methods capture observed behavior, system-level events generally do not specify which effective RBAC permissions are exercised or how those permissions were granted. For example, when a workload communicates with the Kubernetes API server over TLS, connection and data-transfer events alone do not directly reveal the requested API verb, resource, or target object. Process arguments may provide clues about the requested operation, but associating these clues with specific effective permissions and their authorization sources requires RBAC context.

Taken together, these observations highlight the need to establish reliable associations between high-risk effective permissions and runtime behavior, and to leverage the resulting evidence for intrusion detection. Addressing this problem presents three key challenges:

\begin{enumerate}[label=(\arabic*),leftmargin=*,itemsep=3pt,topsep=3pt]

\item \textbf{Lack of Direct Permission-Use Evidence.}
RBAC configurations describe what a subject is authorized to do, but not whether those permissions are actually exercised. System-level events record observed behavior without directly identifying the corresponding permissions, making actual permission use difficult to verify.

\item \textbf{Cross-Modal Semantic Gap.}
Permission subgraphs encode authorization relationships, whereas runtime event sequences capture system-level execution behavior. Distinct permission operations may also produce similar traces, making it challenging to establish accurate cross-modal correspondences.

\item \textbf{Ambiguity Between Legitimate Use and Abuse.}
Legitimate and malicious activities may exercise the same effective permissions. Consequently, evidence of permission use alone is insufficient to identify an intrusion; the associated runtime behavior must also be assessed against the workload's normal execution patterns.

\end{enumerate}

To address these challenges, we propose JANUS, a permission-guided runtime intrusion detection method for Kubernetes. JANUS first extracts high-risk effective permissions and their authorization subgraphs from Kubernetes configurations, and constructs permission--event training pairs through controlled execution and offline Audit verification. A dual-view contrastive model then learns associations between permission subgraphs and runtime behavior to infer permission-use evidence. Finally, JANUS incorporates historical permission evidence into provenance-based anomaly detection through gated fusion, enabling the detection of deviations from normal workload behavior.

We evaluate JANUS across multiple Kubernetes workloads and permission-abuse scenarios. In node-level detection, JANUS achieves an F1 score of 0.3697 and a Matthews correlation coefficient (MCC) of 0.3247, outperforming the evaluated provenance-based detectors in precision, F1, and MCC. Compared with the detector without permission information, JANUS reduces false-positive nodes by approximately 14.3\% while maintaining comparable recall.

Our main contributions are summarized as follows:

\begin{enumerate}[label=(\arabic*),leftmargin=*,itemsep=3pt,topsep=3pt]

\item We develop a method for identifying high-risk effective RBAC permissions and their authorization subgraphs, and construct permission--event training pairs through controlled execution and offline Audit verification.

\item We propose a dual-view cross-modal association model that learns correspondences between permission subgraphs and runtime behavior, enabling the inference of permission-use evidence from system-level observations.

\item We design a permission-conditioned intrusion detection method that incorporates historical permission evidence into provenance-graph representations through temporally constrained gated fusion, allowing anomalous behavior to be assessed in the context of workload permissions.

\item We evaluate JANUS across multiple Kubernetes workloads and permission-abuse scenarios. JANUS achieves an F1 score of 0.3697 and an MCC of 0.3247 in node-level detection, and reduces false-positive nodes by approximately 14.3\% compared with detection without permission conditioning while maintaining comparable recall.

\end{enumerate}

\section{Background and Motivation}

This section provides background on Kubernetes RBAC and runtime provenance, presents a motivating example of permission abuse, and discusses the limitations of static authorization analysis and runtime behavior analysis.

\subsection{Kubernetes RBAC and System Provenance}

\noindent\textbf{Role-Based Access Control.}
Kubernetes workloads run in Pods and typically use ServiceAccount identities to access the API server. RBAC defines permissions through Roles and ClusterRoles, whose rules specify permitted verbs on API resources and subresources within specified API groups. RoleBindings and ClusterRoleBindings associate these roles with subjects. RoleBindings grant permissions within a namespace, even when referencing a ClusterRole, whereas ClusterRoleBindings grant permissions across the cluster. A workload's effective RBAC permissions therefore depend on its ServiceAccount identity, applicable bindings, and associated role rules~\citep{kubernetes_rbac}.

\noindent\textbf{System Provenance.}
System-call monitoring captures process, file, and network activity performed by workloads, which can be represented as provenance graphs~\citep{kairos,orthrus}. In these graphs, nodes represent system entities such as processes, files, and network endpoints, while directed, timestamped edges capture their interactions. Whereas RBAC specifies authorized operations, provenance graphs capture observed execution and its context.

\subsection{Motivating Example}

\begin{figure*}[t]
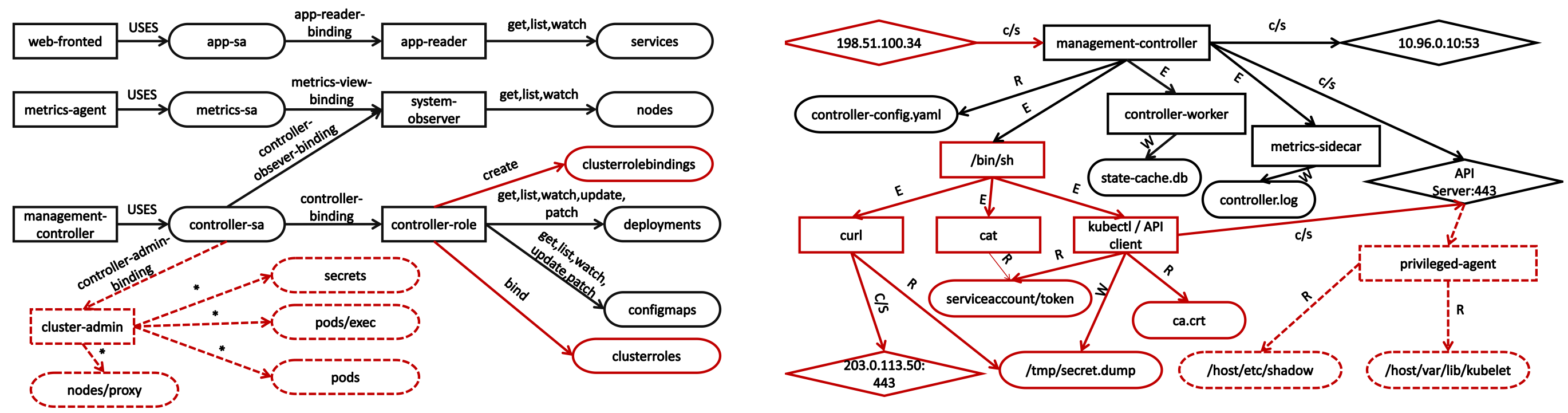

\centering
\captionsetup{skip=3pt}
\begin{minipage}[t]{0.49\textwidth}
\vspace{0pt}
\centering\includegraphics[width=\linewidth,trim=0 37.2bp 0 50.4bp,clip]{1a.png}
\par\smallskip{\small (a) Static permission-reachability graph}
\end{minipage}\hfill
\begin{minipage}[t]{0.49\textwidth}
\vspace{0pt}
\centering\includegraphics[width=\linewidth,trim=0 37.2bp 0 50.4bp,clip]{1b.png}
\par\smallskip{\small (b) Runtime provenance graph}
\end{minipage}
\caption{Motivating example of permission abuse in Kubernetes. }
\label{fig:motivation}
\end{figure*}

Consider a Kubernetes cluster in which a third-party management controller accesses the API server using the ServiceAccount \texttt{controller-sa}. As illustrated in Fig.~\ref{fig:motivation}, an attacker exploits a remote-code-execution vulnerability in the controller, retrieves its mounted ServiceAccount token, and uses the controller's existing RBAC permissions to create a ClusterRoleBinding that grants the cluster-admin role. With the resulting privileges, the attacker reads sensitive Secrets and, under the scenario's permissive admission and network settings, deploys a privileged workload with host-directory mounts to collect and exfiltrate data to an external endpoint.

Figure~\ref{fig:motivation}(a) shows the authorization relationships that enable this escalation. The existing ClusterRoleBinding \path{controller-binding} associates \path{controller-sa} with the ClusterRole \path{controller-role}. Its rules grant \texttt{create} on \texttt{clusterrolebindings} and \texttt{bind} on \texttt{clusterroles}, with the latter restricted to the ClusterRole named \path{cluster-admin}. Together, these permissions allow the attacker to create \path{controller-admin-binding}, granting \path{controller-sa} cluster-administrator privileges. The initial configuration reveals a potential escalation path but cannot establish whether it is exercised, whereas a subsequent configuration snapshot reflects the newly created binding and the resulting privilege expansion.

Figure~\ref{fig:motivation}(b) shows the runtime provenance graph constructed from Sysdig events, which capture container-attributed interactions among processes, files, and network endpoints. During normal operation, the controller reads configuration files, starts worker processes, and communicates with the API server to manage cluster resources. Following the compromise, the controller spawns \texttt{/bin/sh}, which invokes \texttt{cat} to read the ServiceAccount token and an API client to communicate with the API server over TLS. The subsequently created \path{privileged-agent} accesses sensitive host files, Kubelet directories, and the container-runtime socket, while \texttt{curl} reads staged data and sends it to an external endpoint. However, legitimate and malicious API requests may appear as similar encrypted connections to the same API server. Such network events do not directly reveal which RBAC permissions were exercised.

\subsection{Limitations}

The motivating example highlights two Limitations in detecting Kubernetes permission abuse from authorization and runtime evidence.

\noindent\textbf{Permission Reachability Does Not Imply Use.}
Existing approaches analyze Kubernetes RBAC configurations to identify risky permissions and potential privilege-escalation paths. EPScan further combines RBAC configuration analysis with static program analysis to identify excessive permissions in third-party applications~\citep{epscan}. While these techniques reveal security risks associated with granted permissions, they do not establish whether a particular permission is exercised during execution.

In Fig.~\ref{fig:motivation}(a), the authorization path associated with \path{controller-sa} permits the creation of a ClusterRoleBinding that grants \path{cluster-admin} privileges. However, this path exists regardless of whether the permission is unused, exercised legitimately, or abused. Although the subsequent creation of \path{controller-admin-binding} changes the RBAC configuration, the updated state alone does not reveal the runtime behavior behind the change. Moreover, reading sensitive Secrets through existing permissions leaves the authorization state unchanged. Consequently, treating all reachable high-risk permissions as suspicious may produce false positives, while monitoring only RBAC changes may miss attacks that do not modify authorization relationships.

\begin{observationbox}[before skip=1pt,top=2pt,bottom=2pt]
\noindent\textbf{Observation 1.}
Static RBAC analysis identifies high-risk capabilities and authorization changes, but cannot reliably associate permission use with observed execution. Runtime evidence is needed to establish when and how these permissions are exercised.
\end{observationbox}

\noindent\textbf{Provenance Lacks RBAC Context.}\,\mbox{Provenance-based} intrusion detection systems, such as KAIROS and ORTHRUS, model system execution to identify anomalous behavior~\citep{kairos,orthrus}. Although these approaches capture execution dependencies and behavioral deviations, system-level provenance does not directly expose the Kubernetes RBAC permissions underlying API requests.

As shown in Fig.~\ref{fig:motivation}(b), both legitimate controller operations and malicious API requests may appear as encrypted connections to the same API server. The provenance graph captures the surrounding process, file, and network activity, but does not directly reveal the API operations or the authorization paths involved. Kubernetes Audit logs can provide request-level information and authorization decisions, but these records do not by themselves connect API requests to the broader system-level execution context. Without such associations, provenance-based detection lacks the permission context needed to interpret observed behavior in relation to a workload's effective privileges.

\begin{observationbox}
\noindent\textbf{Observation 2.}
Runtime provenance captures execution behavior but does not directly link it to effective RBAC permissions and their authorization sources. Associating runtime evidence with static authorization context is necessary for permission-aware intrusion detection.
\end{observationbox}

\section{Threat Model}

We consider an attacker who gains code execution within a Kubernetes workload by exploiting an application vulnerability or introducing a malicious dependency. Once the workload is compromised, the attacker can access its available ServiceAccount credentials and issue Kubernetes API requests using the associated effective RBAC permissions. The attacker may abuse these permissions to access sensitive resources or escalate privileges by modifying RBAC bindings when permitted by Kubernetes authorization.

Following prior provenance-based intrusion detection systems~\citep{kairos,orthrus}, we assume the integrity of the underlying monitoring infrastructure. We further assume that Kubernetes authorization data can be collected reliably. Specifically, we make the following assumptions:

\begin{enumerate}[label=(\arabic*),leftmargin=*,itemsep=3pt,topsep=3pt]

\item \textbf{Trusted authorization data.}
RBAC configuration snapshots accurately reflect the authorization state at collection time. Kubernetes Audit records used for offline verification faithfully record the relevant API operations.

\item \textbf{Trusted runtime events.}
The Sysdig event collection mechanism is protected from attacker interference, and the integrity of collected events is preserved during transmission and storage.

\item \textbf{Infrastructure integrity.}
The host kernel and Kubernetes control-plane components remain uncompromised. Attacker-initiated authorization changes through legitimate Kubernetes API requests are within scope, whereas attacks that compromise the trusted infrastructure are excluded.

\end{enumerate}

We also assume that the detector and its training pipeline are trusted, excluding data and model poisoning attacks. Although adaptive evasion against provenance-based detectors has been studied~\citep{provninja}, evaluating such attacks is beyond the scope of this work.

\section{JANUS Framework}

\subsection{Overview}

\begin{figure*}[t]
\centering
\includegraphics[width=\textwidth]{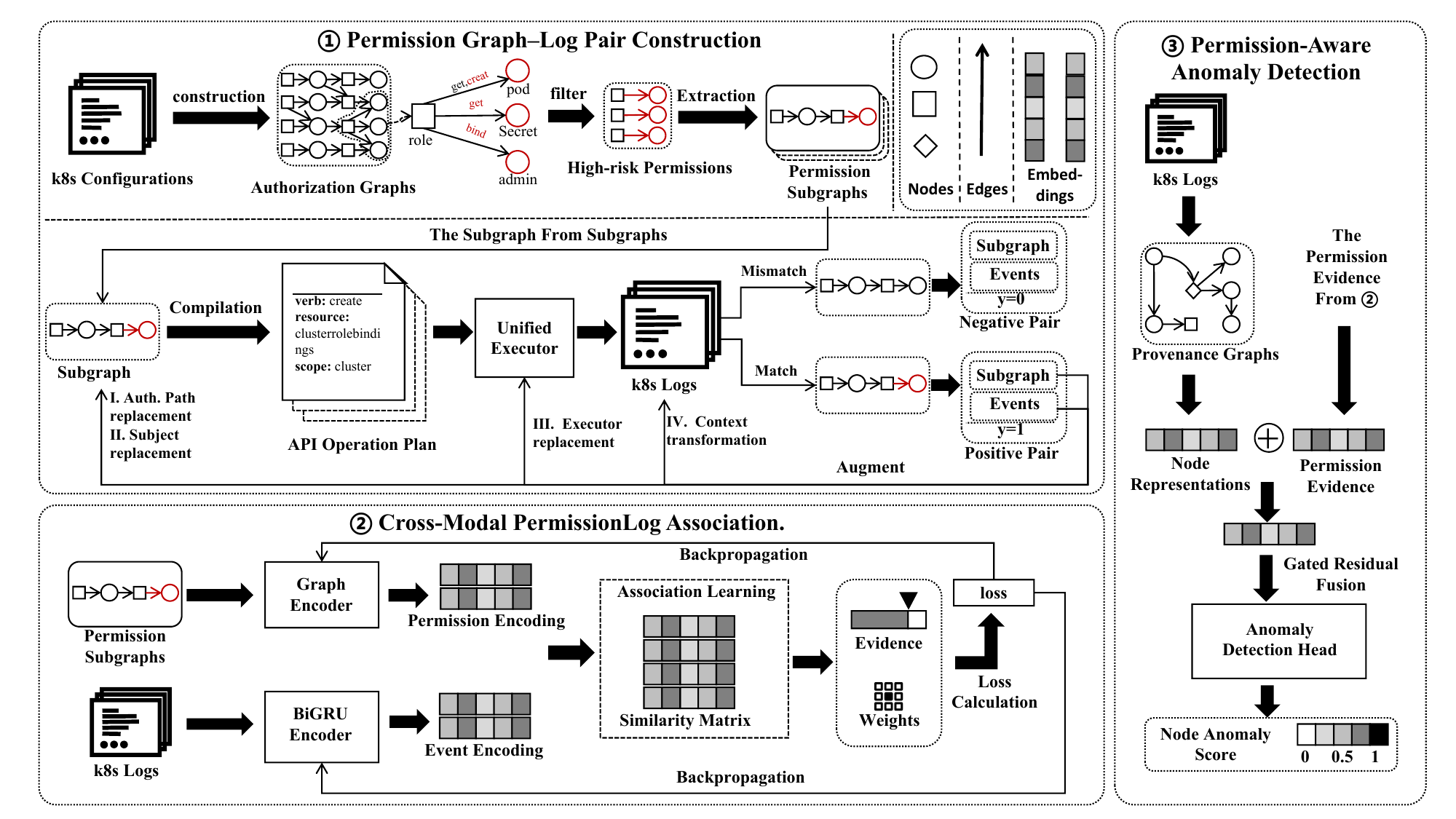}
\caption{Overview of JANUS.}
\label{fig:overview}
\end{figure*}

JANUS integrates effective Kubernetes RBAC permissions with runtime provenance to support permission-aware, node-level intrusion detection. Figure~\ref{fig:overview} presents the overall architecture, which comprises three main components.

\noindent\textbf{\textcircled{\scriptsize 1} Permission Graph--Log Pair Construction}
We identify high-risk effective permissions and extract their authorization subgraphs from Kubernetes RBAC configurations. Controlled API execution and Sysdig event collection, verified against Kubernetes Audit records, establish positive permission--event pairs. Permission-preserving augmentation and semantic hard-negative sampling further enrich the training data.

\noindent\textbf{\textcircled{\scriptsize 2} Cross-Modal Permission--Log Association.}
We train a dual-view model to align permission subgraphs with workload-specific runtime event windows through contrastive learning and matching supervision. The model produces window-level association scores to infer permission use, while local semantic matching provides fine-grained event-support weights.

\noindent\textbf{\textcircled{\scriptsize 3} Permission-Aware Anomaly Detection.}
We incorporate historical permission evidence into runtime provenance-graph representations through temporally constrained gated fusion. Trained on benign execution, the detector identifies anomalous nodes by measuring deviations from normal workload behavior in the context of associated permissions.

\subsection{Permission Graph--Log Pair Construction}
\label{sec:pair-construction}

To learn associations between effective permissions and runtime behavior, we construct labeled pairs of high-risk permission subgraphs and runtime event windows. High-risk permission analysis identifies the authorization targets, verified API executions establish positive correspondences, and semantically mismatched operations provide negative supervision. Each pair is defined as:
\begin{equation}\label{eq:training-pair}
d_{ik}=\langle G_i^H,B_k,y_{ik}\rangle,
\qquad y_{ik}\in\{0,1\},
\end{equation}
where $G_i^H$ denotes a high-risk permission subgraph, $B_k$ represents a runtime event window, and $y_{ik}$ indicates whether the target permission matches the verified API operation associated with that window.

\noindent\textbf{High-risk effective permission analysis.}
Individual RBAC rules do not fully characterize a workload's effective permissions without its identity, role bindings, and authorization scope. We therefore construct a permission graph $G_t^P=(V_t^P,E_t^P)$ from a Kubernetes configuration snapshot $K_t$ at time $t$. The graph contains four node types: Workload, ServiceAccount, Role/ClusterRole, and API Target. Three relations, \textsc{USES}, \textsc{BOUND\_TO}, and \textsc{ALLOWS}, capture identity use, role binding, and resource authorization, respectively, with binding information retained on the corresponding relations.

We normalize the RBAC operations associated with these authorization paths into atomic permissions:
\begin{equation}\label{eq:atomic-permission}
h_i=\langle v_i,g_i,r_i,u_i,n_i\rangle,
\end{equation}
where $v_i$, $g_i$, $r_i$, $u_i$, and $n_i$ denote the verb, API group, resource, subresource, and resource-name constraints, respectively. Resolving the applicable bindings and authorization scope yields an effective permission:
\begin{equation}\label{eq:effective-permission}
\eta_i=\langle w_i,s_i,h_i,\ell_i,\mathcal P_i\rangle,
\end{equation}
where $w_i$ and $s_i$ identify the workload and its ServiceAccount, $\ell_i$ denotes the effective authorization scope, and $\mathcal P_i$ contains the authorization paths. The resolution accounts for Kubernetes RBAC semantics, including scope, wildcards, and applicable resource-name constraints~\citep{kubernetes_rbac}.

We identify high-risk effective permissions using a catalog $\mathcal C_H$ based on Kubernetes RBAC semantics and documented security risks~\citep{kubernetes_rbac,kubernetes_rbac_good_practices}. We organize these permissions into 13 families, as summarized in Table~\ref{tab:highrisk-families}, and extract their authorization subgraphs $G_i^H=(V_i^H,E_i^H,h_i)$.

% Representative resources; additional family members are described in the text.
\begin{table*}[t]
\centering
\begingroup
\papertablestyle{3pt}
\begin{tabularx}{\textwidth}{@{}L{0.18\textwidth}L{0.30\textwidth}L{0.17\textwidth}Y@{}}
\toprule
\rowcolor{TableHeaderBg}
\textbf{Permission family} & \textbf{Representative resource} & \textbf{Verbs} & \textbf{Potential capability} \\
\midrule
Secret access & \texttt{secrets} & \texttt{get/list/watch} & Read data and credentials \\
ServiceAccount tokens & \texttt{serviceaccounts/token} & \texttt{create} & Request identity tokens \\
CSR operations & \texttt{certificatesigningrequests} & \texttt{create} & Request certificates \\
Pod remote interaction & \texttt{pods/exec} & \texttt{create/get} & Execute remote commands \\
Workload mutation & \texttt{pods, deployments, jobs} & \texttt{create/update/patch} & Change execution configuration \\
Volume control & \texttt{persistentvolumes} & \texttt{create/update/patch} & Control storage definitions \\
Role-rule expansion & \texttt{roles, clusterroles} & \texttt{create/update/patch} & Expand role permissions \\
Binding expansion & \texttt{rolebindings, clusterrolebindings} & \texttt{create/update/patch} & Assign permissions to subjects \\
Impersonation & \texttt{users, groups, serviceaccounts} & \texttt{impersonate} & Act as another identity \\
Kubelet proxy & \texttt{nodes/proxy} & \texttt{get/create} & Access Kubelet APIs \\
Node modification & \texttt{nodes} & \texttt{update/patch} & Modify node settings \\
Webhook control & \texttt{validatingwebhookconfigurations} & \texttt{create/update/patch} & Alter admission behavior \\
Namespace modification & \texttt{namespaces} & \texttt{update/patch} & Modify namespace settings \\
\bottomrule
\end{tabularx}
\endgroup
\caption{High-risk Kubernetes permission families.}
\label{tab:highrisk-families}
\end{table*}

\noindent\textbf{Positive Pair Construction.}
Authorization subgraphs characterize high-risk capabilities but do not establish actual permission use. We therefore execute matching Kubernetes API operations under the corresponding ServiceAccounts in a controlled environment. Kubernetes Audit records provide the requesting subject, operation, resource, and recorded outcome~\citep{kubernetes_audit}, while Sysdig captures the associated system-level events~\citep{sysdig}.

We associate verified API operations with Sysdig events through process attribution, temporal alignment, and observable execution cues, obtaining windows $B_k$ of at most 32 events. A pair $(G_i^H,B_k)$ receives $y_{ik}=1$ only when the verified operation matches the target effective permission, including its applicable authorization constraints, and the window contains attributable execution evidence. Audit records are used exclusively for offline verification and are excluded from model inputs.

Positive labels must reflect the specific permission operation rather than related requests or inferred outcomes. For example, creating a binding or modifying a role does not necessarily establish the use of \texttt{bind} or \texttt{escalate}, respectively~\citep{kubernetes_rbac}. We therefore validate permission-specific authorization semantics and avoid assigning labels based on unverified downstream effects.

Verified executions provide reliable supervision, but the resulting traces may contain execution-specific cues unrelated to permission semantics. We address this variability through permission-preserving augmentation across authorization structures and runtime execution contexts. On the authorization side, we vary ServiceAccount identities and binding paths while retaining the target permission semantics. On the runtime side, we vary API clients and benign background activity to produce different traces of the corresponding operation.

We constrain these transformations using a normalized permission signature:
\begin{equation}\label{eq:permission-signature}
\operatorname{sig}(G_i^H)=
\langle v_i,g_i,r_i,u_i,n_i,
\kappa(\ell_i),\mathcal R_i,\mathcal S_i\rangle,
\end{equation}
where $\kappa(\ell_i)$ denotes the authorization scope type, and $\mathcal R_i$ and $\mathcal S_i$ represent risk-type and target labels, respectively. Each augmentation transformation $\zeta$ must satisfy
\begin{equation}\label{eq:semantic-preservation}
\operatorname{sig}(G_i^{H,\zeta})
=
\operatorname{sig}(G_i^H).
\end{equation}
Since signature equality alone does not guarantee equivalent effective authorization, augmented instances are retained only after re-execution, verification of the applicable authorization constraints and API operation, and runtime event alignment.

\noindent\textbf{Negative Pair Construction.}
Permission-preserving augmentation diversifies matching permission--event pairs, but distinguishing semantically different permissions with similar runtime traces still requires negative supervision. We construct ordinary negatives from verified windows whose anchor operations differ from the target permission in semantics. Hard negatives deliberately preserve either the resource or the verb while changing the other, creating semantically close mismatches.

Candidates with equivalent permission semantics, insufficient verification, or unreliable event alignment are excluded. The remaining pairs receive $y_{ik}=0$, indicating a mismatch between the target permission and the verified anchor operation, rather than proving that the target permission was never exercised within the window.

\subsection{Cross-Modal Permission--Log Association}
\label{sec:association}

The verified pairs constructed in Section~\ref{sec:pair-construction} provide supervision for permission--event associations, but permission subgraphs and runtime event windows differ in structure and semantics. We therefore encode the two modalities separately and learn their associations by contrasting matching and mismatching pairs. The model produces window-level association scores and event-level support weights for subsequent intrusion detection.

\noindent\textbf{Dual-View Encoding.}
Window-level association and fine-grained event matching require both global and local representations. We use a relation-aware graph encoder and a sequence encoder to obtain complementary representations in a common $d$-dimensional space. Relation-aware message passing provides a general mechanism for encoding typed graph relations~\citep{rgcn}.

For each permission subgraph $G_i^H$, we select a single authorization path consistent with the target subject and operation, forming a four-node graph comprising Workload, ServiceAccount, Role/ClusterRole, and API Target. Node features combine entity-type and permission-semantic token embeddings. Two layers of relation-aware message passing capture the authorization structure, followed by mean pooling, linear projection, and $\ell_2$ normalization to obtain the global permission representation $p_i\in\mathbb R^d$. At inference, identical encoder inputs can share a cached representation, while their authorization sources are retained separately for attribution.

The pooled permission representation supports global association but does not expose individual permission attributes separately for event-level matching. We therefore introduce six local semantic units corresponding to the verb, API group, resource, subresource, scope, and risk type. Shared token embeddings followed by $\ell_2$ normalization produce $U_i=[u_{i1},\ldots,u_{i6}]\in\mathbb R^{6\times d}$. These units bypass graph message passing and provide attribute-level representations for fine-grained event matching.

The runtime encoder likewise provides global and event-level representations. For each window $B_k$, we embed event type, process category, execution result, and command semantics together with positional information. A bidirectional GRU~\citep{gru} captures temporal dependencies and produces event representations $b_{kj}$, whose $\ell_2$-normalized forms $z_{kj}$ constitute $Z_k\in\mathbb R^{L_k\times d}$, where $L_k$ denotes the window length. Attention pooling, linear projection, and $\ell_2$ normalization yield the global event representation $v_{kc}$ for chunk $c$. The encoder supports non-overlapping chunks of at most 32 events, with each training window forming a single chunk.

\noindent\textbf{Contrastive Learning.}
Permission-preserving augmentation produces multiple valid runtime manifestations of the same target permission semantics, while distinct permissions may generate similar event sequences. We therefore combine multi-positive contrastive learning, binary matching, and hard-negative ranking to accommodate execution variability and discriminate mismatched operations.

Using the normalized global representations, we define the temperature-scaled permission--window similarity as $s_{ik}=\max_c(p_i^\top v_{kc})/T$, where $T$ is the temperature parameter. Learnable scaling and shifting yield the matching logit $\ell_{ik}=\alpha s_{ik}+\beta$, with $q_{ik}^{\mathrm{train}}=\sigma(\ell_{ik})$ providing a sigmoid matching score for binary supervision.

Verified positive pairs sharing the same target permission signature are grouped as multiple positives. For a batch of $b$ positive pairs, we compute all in-batch permission--window scores. Let $\mathcal I$ and $\mathcal B$ denote the permission and window index sets, respectively, and let $P(i)$ contain the verified positive windows grouped with permission $i$. The permission-to-event loss is
\begin{equation}\label{eq:multi-positive-contrastive}
\mathcal L_{p\rightarrow e}
=-\frac{1}{|\mathcal I|}
\sum_{i\in\mathcal I}
\log
\frac{\sum_{k\in P(i)}\exp(s_{ik})}
{\sum_{k\in\mathcal B}\exp(s_{ik})}.
\end{equation}
Exchanging the two modalities yields $\mathcal L_{e\rightarrow p}$, and their average defines $\mathcal L_{\mathrm{multi}}$. Unlike the standard supervised contrastive objective~\citep{supcon}, which averages individual positive log probabilities, this formulation maximizes the aggregate softmax probability mass assigned to positive windows.

The contrastive objective captures relative associations within a batch, while binary matching and hard-negative ranking provide complementary pairwise constraints. We apply binary cross-entropy $\mathcal L_{\mathrm{BCE}}$ to balanced positive and negative pairs using $q_{ik}^{\mathrm{train}}$. A margin-ranking loss $\mathcal L_{\mathrm{rank}}$ encourages the unscaled maximum cosine similarity of a positive pair, $c_i^+$, to exceed that of its corresponding hard-negative pair, $c_i^-$. The hinge loss uses a margin of $0.15$ and is averaged over valid hard negatives, contributing zero when none are available. The overall objective is
\begin{equation}\label{eq:association-objective}
\mathcal L_{\mathrm{assoc}}
=
\mathcal L_{\mathrm{multi}}
+
\mathcal L_{\mathrm{BCE}}
+
\mathcal L_{\mathrm{rank}}.
\end{equation}

\noindent\textbf{Association Inference.}
Window-level association scores measure permission--event correspondence but do not identify the events contributing to each match. We complement these scores with local semantic matching to obtain event-level support weights for subsequent detection.

For event $j$ in window $B_k$, we compute its local semantic similarity to permission $i$ as $t_{ikj}=\max_m u_{im}^{\top}z_{kj}$, taking the maximum over the six permission units. The event-support weight is
\begin{equation}\label{eq:event-support}
a_{ikj}=
\frac{\exp(t_{ikj})}
{\sum_{j'\in\operatorname{Valid}(B_k)}
\exp(t_{ikj'})},
\end{equation}
where $\operatorname{Valid}(B_k)$ indexes non-padding events. The resulting weight measures relative event support within the window rather than independently establishing permission use. It is derived from representations learned with pair-level supervision, without requiring event-level labels.

During online inference, each process maintains a rolling event window, and permission associations are computed as new events arrive. For each retained candidate permission, we apply the learned matching function to obtain the current window-level score $q_{ik}^{\mathrm{deploy}}$, and cache it together with the last-event support weight $a_{ikL_k}$ and permission representation $p_i$ for subsequent permission-aware anomaly detection. Previously processed events are not updated retrospectively.

\subsection{Permission-Aware Anomaly Detection}

\label{sec:permission-detection}

Evidence of permission use alone does not indicate malicious behavior, as legitimate and malicious activities may exercise the same high-risk permissions. We therefore incorporate temporally aligned historical permission evidence into runtime provenance representations and learn normal permission-conditioned behavior from benign executions. Deviations from the learned behavior are scored at the event level and aggregated to identify anomalous nodes.

\noindent\textbf{Permission Evidence Fusion.}
Permission evidence can inform anomaly detection only if it precedes the event being predicted; otherwise, the detector may exploit information derived from the target event itself. For an event $e_j=(u,v)$ at time $t_j$, a two-layer GraphTransformer encodes the provenance history available before the current batch to obtain source and destination representations $h_u$ and $h_v$. Historical permission evidence then conditions the source representation, while the destination representation remains unchanged.

For each previously observed event $e_\ell$, let $q_{i\ell}$ and $a_{i\ell}$ denote the deployment association score and last-event support weight for candidate permission $i$, respectively, as defined in Section~\ref{sec:association}. We aggregate the associated permission representations as $r_\ell=\sum_i q_{i\ell}a_{i\ell}p_i$. This evidence is stored with the executing process, with process-creation events attributed to the newly created child. When predicting $e_j$, we retrieve only the source process's most recent valid evidence preceding $t_j$, denoted $r_{u,<t_j}$.

The relevance of historical permission evidence depends on the current runtime context. We therefore use gated residual fusion to regulate its contribution to the source representation. The evidence is first projected into the node representation space as $c_j=\operatorname{Proj}(r_{u,<t_j})$, followed by
\begin{equation}\label{eq:gated-fusion}
\begin{aligned}
g_j &= \sigma\left(W_g[h_u\Vert c_j]+b_g\right),\\
\widetilde h_u &= h_u+g_j\odot c_j,
\end{aligned}
\end{equation}
where $W_g$ and $b_g$ are learnable parameters, $\Vert$ denotes concatenation, and $\odot$ denotes element-wise multiplication. When no valid historical evidence is available, we set $r_{u,<t_j}=0$. The bias-free projection then yields $c_j=0$, preserving the original representation $\widetilde h_u=h_u$.

Permission evidence is valid only if it belongs to the same authorization snapshot and was generated within one second before the predicted event. Each process retains its most recent valid association evidence; events without retained associations do not overwrite the cached result, which remains subject to the same validity constraints. To prevent information leakage, graph history and permission evidence are updated only after all events in the current batch have been predicted. The fused representation is used only for current-event decoding and does not modify historical graph features.

\noindent\textbf{Anomaly Scoring.}
With historical permission evidence incorporated into the source representation, the detector learns normal event-type patterns from separate benign runtime data while keeping the association model frozen. For each event $e_j$, the detector predicts its type from the fused source representation $\widetilde h_u$ and destination representation $h_v$. The negative log-probability of the observed event type defines the anomaly score:
\begin{equation}\label{eq:event-anomaly}
A(e_j)=-\log P_\theta
\left(\tau_j\mid\widetilde h_u,h_v\right),
\end{equation}
where $\tau_j$ denotes the observed event type and $\theta$ represents the detector parameters. Higher scores indicate events that are less consistent with the learned benign patterns.

Event-level scores are aggregated to identify anomalous nodes. The base node score is $S_{\mathrm{base}}(v)=\max_{e_j\in\operatorname{Inc}(v)}A(e_j)$, where $\operatorname{Inc}(v)$ denotes the incident scored events. During chronological scoring, a process-creation event also propagates the parent's score accumulated up to that event to the child if it exceeds the child's current score. Later changes to the parent's score are not propagated retrospectively. Nodes whose final scores $S(v)$ exceed the 85th percentile of benign validation node scores are classified as anomalous.

\section{Evaluation}

We evaluate JANUS in terms of permission--event association, node-level intrusion detection, component contributions, hyperparameter sensitivity, and computational efficiency. Specifically, we address the following research questions:

\noindent\textbf{RQ1:}
How accurately does JANUS associate permissions with runtime events, and how does its node-level intrusion detection compare with existing approaches?

\noindent\textbf{RQ2:}
How do the key components of JANUS contribute to permission--event association and intrusion detection?

\noindent\textbf{RQ3:}
How do key hyperparameters affect association and detection performance?

\noindent\textbf{RQ4:}
What are the computational costs of permission--event association and permission-aware anomaly detection?

\subsection{Experimental Setup}

\noindent\textbf{Datasets.}
We obtain permission information from Kubernetes configuration snapshots, collect runtime events using Sysdig, and use Kubernetes Audit records to verify controlled API executions. Separate datasets are constructed for permission--event association and runtime intrusion detection.

Table~\ref{tab:dataset-statistics} summarizes the association dataset, comprising 1,872 event windows from 936 controlled execution groups. The data are divided into training, validation, and test sets, all covering 13 permission queries. Training samples are collected through kubectl executions, while validation and test samples use two distinct HTTPS execution contexts. No raw windows or identical encoded event sequences are shared across the three sets.

For runtime intrusion detection, collection rounds 1--10, 11--13, and 14--18 form the training, validation, and test sets, respectively. The detector is trained exclusively on benign events. The test set contains 1,612,875 raw events, with 1,535 malicious and 18,960 benign nodes included in the labeled evaluation. Node labels are determined from attack execution records and direct causal associations; nodes without established labels are excluded.

\begin{table}[!htbp]
\centering
\begingroup
\papertablestyle{3pt}
\begin{tabularx}{\columnwidth}{@{}YRRR@{}}
\toprule
\rowcolor{TableHeaderBg}
\textbf{Data} & \textbf{Training} & \textbf{Validation} & \textbf{Test} \\
\midrule
Event windows & 936 & 468 & 468 \\
Positive pairs & 468 & 234 & 234 \\
Negative pairs & 11,700 & 5,850 & 5,850 \\
\bottomrule
\end{tabularx}
\endgroup
\caption{Association dataset statistics.}
\label{tab:dataset-statistics}
\end{table}

\noindent\textbf{Experimental Environment.}
JANUS uses CPU-based execution for permission--event association and node-level intrusion detection. The feature-disjoint association experiments and efficiency measurements used Python 3.11.15 and PyTorch 2.9.1 (CPU build). The efficiency experiments were conducted on a server equipped with two Intel Xeon Silver 4114 processors (2.20 GHz) and 125 GiB of system memory, running Linux kernel 5.15.0-139-generic. The efficiency experiments used PyTorch Geometric 2.7.0.

% TODO: Add the original training and evaluation hardware/software
% environment, including CPU, GPU, memory, and software versions.
% The supplementary export environment is not a substitute.

\noindent\textbf{Metrics.}
For permission--event association, Recall@1, Recall@5, Recall@10, and mean reciprocal rank (MRR) measure retrieval performance, while Precision, Recall, and F1 evaluate binary matching. Node-level detection is evaluated using Precision, Recall, F1, AUROC, AUPRC, false positive rate (FPR), and Matthews correlation coefficient (MCC), with malicious nodes as the positive class. Precision--recall analysis complements ROC analysis under class imbalance~\citep{pr_imbalanced}.

% TODO: Specify the retrieval query direction, candidate set,
% and ranking protocol for Recall@K and MRR.

\noindent\textbf{Parameter Settings.}
The association embedding dimension is 128, and the contrastive temperature is 0.12. Each batch contains 24 positive and 24 negative pairs, with a target ratio of 12:4:8 for hard, control, and other ordinary negatives. Control negatives correspond to the dataset category \texttt{granted\_but\_not\_used}, representing available permissions that do not match the verified operations anchoring the sampled windows. If a designated negative pool is insufficient, the remaining slots are filled from the ordinary negative pool or, if necessary, the full negative pool. The association rejection threshold is selected using the validation set.

The detector uses a two-layer GraphTransformer. The first layer has eight attention heads, each with 16 dimensions, and the second produces 128-dimensional node representations. The default neighbor sample size is 20, and the event replay batch size is 256. For the complete model and each ablation, the detection threshold is set to the 85th percentile of the corresponding benign validation node-score distribution, whereas the static analysis uses the 99th percentile.

\subsection{RQ1: Overall Effectiveness}

\noindent\textbf{Setup.}
We evaluate permission--event association, node-level intrusion detection, and static RBAC analysis. For permission--event association, we compare JANUS with TF-IDF cosine similarity and BM25~\citep{bm25} using 13 permission queries and 468 candidate test windows. Lexical statistics are computed from the training set. Recall@$k$ measures the fraction of relevant windows retrieved per query, averaged across queries. Mean reciprocal rank (MRR) is computed from the first relevant result for each query, while average precision (AP) is calculated across all permission--window pairs.

\begin{table*}[!t]
\centering
\begingroup
\papertablestyle{4pt}
\begin{tabularx}{\textwidth}{@{}l*{8}{R}@{}}
\toprule
\rowcolor{TableHeaderBg}
\textbf{Method} & \textbf{Prec.} & \textbf{Rec.} & \textbf{F1} & \textbf{R@1} & \textbf{R@5} & \textbf{R@10} & \textbf{MRR} & \textbf{AP} \\
\midrule
TF-IDF & 0.7500 & \textbf{0.9231} & 0.8276 & 0.0513 & 0.2564 & 0.5128 & 0.9271 & 0.8650 \\
BM25 & 0.8745 & \textbf{0.9231} & 0.8981 & \textbf{0.0556} & \textbf{0.2778} & \textbf{0.5513} & \textbf{1.0000} & \textbf{0.9582} \\
\textbf{JANUS} & \textbf{1.0000} & 0.9188 & \textbf{0.9577} & 0.0513 & 0.2564 & 0.5128 & 0.9234 & 0.9268 \\
\bottomrule
\end{tabularx}
\endgroup
\caption{Permission--event association results on the test set.}
\label{tab:association-comparison}
\end{table*}

For binary matching, all three methods select their decision thresholds by maximizing F1 on the validation set, with ties resolved in favor of the higher threshold. The selected thresholds are fixed for evaluation on the independent test set. JANUS uses the checkpoint from the final training epoch, without selecting model parameters based on test performance.

For node-level intrusion detection, we compare JANUS with THREATRACE~\citep{threatrace}, MAGIC~\citep{magic}, KAIROS~\citep{kairos}, FLASH~\citep{flash}, and ORTHRUS~\citep{orthrus}. All methods use the same data split and are evaluated on 20,495 labeled test nodes. ORTHRUS retains its original scoring mechanism.

For static RBAC analysis, we evaluate KubeLinter~\citep{kube_linter} using built-in RBAC checks and Kubescape~\citep{kubescape} using NSA and MITRE rules. Configuration findings are mapped to corresponding processes, with an alert threshold set at the 99th percentile of benign validation scores. The evaluation includes 4,338 processes, comprising 4,248 malicious and 90 benign processes.

\noindent\textbf{Results and Analysis.}
Table~\ref{tab:association-comparison} presents the permission--event association results. JANUS achieves the highest binary matching F1 of 0.9577, outperforming BM25 (0.8981) and TF-IDF (0.8276). Compared with BM25, JANUS identifies 215 positive pairs versus 216, while reducing false-positive matches from 31 to zero. This result highlights the selectivity of learned permission--event associations, which distinguish authorized behavior from unrelated runtime events with few missed positive matches. BM25 achieves higher retrieval ranking scores, indicating that JANUS's advantage lies in binary matching at the selected thresholds rather than retrieval ranking.

\begin{table*}[!t]
\centering
\begingroup
\papertablestyle{5pt}
\begin{tabularx}{\textwidth}{@{}l*{6}{R}@{}}
\toprule
\rowcolor{TableHeaderBg}
\textbf{Method}
& \metrichead{Precision}
& \metrichead{Recall}
& \metrichead{F1}
& \metrichead{AUPRC}
& \metrichead{FPR}
& \metrichead{MCC} \\
\midrule
THREATRACE
& 0.1554 & 0.4450 & 0.2304
& \textbf{0.2345} & 0.1958 & 0.1598 \\
MAGIC
& 0.1017 & 0.0664 & 0.0804
& 0.0857 & \textbf{0.0475} & 0.0231 \\
KAIROS
& 0.2105 & \textbf{0.6267} & 0.3151
& 0.1572 & 0.1903 & 0.2759 \\
FLASH
& 0.0759 & 0.2730 & 0.1188
& 0.0763 & 0.2689 & 0.0024 \\
ORTHRUS
& 0.2312 & 0.4306 & 0.3009
& 0.1512 & 0.1159 & 0.2391 \\
\textbf{JANUS}
& \textbf{0.2733}
& 0.5713
& \textbf{0.3697}
& 0.1963
& 0.1230
& \textbf{0.3247} \\
\bottomrule
\end{tabularx}
\endgroup
\caption{Comparison of node-level intrusion detection methods.}
\label{tab:rq1-runtime}
\end{table*}

We further evaluate JANUS against five provenance-based intrusion detectors at the node level. As shown in Table~\ref{tab:rq1-runtime}, JANUS achieves the highest precision, F1, and MCC, improving F1 by 22.9\% and 17.3\% over ORTHRUS and KAIROS, respectively.

\begin{figure}[!htbp]
\centering
\includegraphics[width=\linewidth]{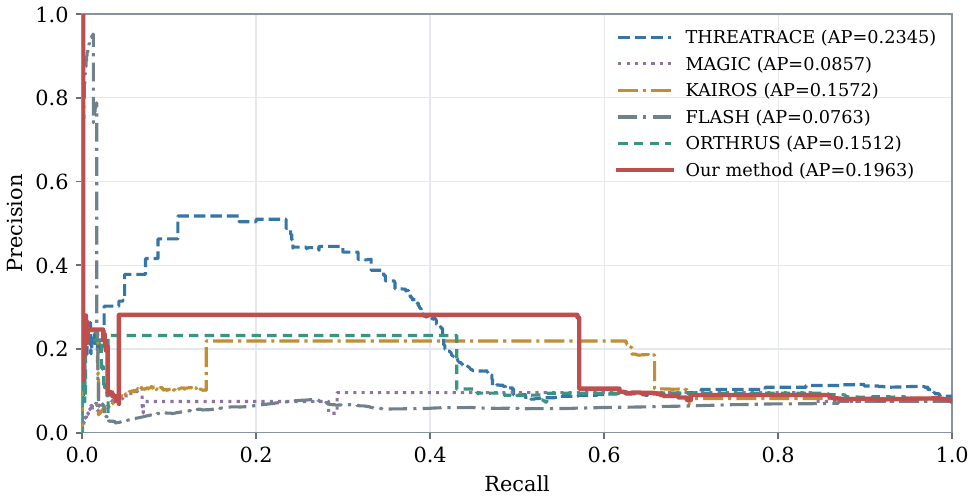}
\caption{Precision--recall curves for node-level intrusion detection.}
\label{fig:rq1-pr-curve}
\end{figure}

The improvement is most pronounced in attack-associated process detection. Compared with ORTHRUS, JANUS nearly doubles process recall, from 26.96\% to 52.96\%. Examining the additional detections reveals that 214 of the 220 malicious nodes detected by JANUS but missed by ORTHRUS cross the detection threshold only after causal score inheritance at process creation. This allows anomaly evidence from an ancestor process to contribute to the detection of its descendants, even when their own incident-event scores are insufficient to trigger an alert. The resulting increase in process coverage is particularly relevant to attacks that involve multiple generations of processes, such as privileged workload creation.

JANUS also achieves more selective detection of resource entities. Compared with ORTHRUS, it reduces resource-node false positives by 73.2\% while retaining the same set of 30 detected malicious resources. Unlike the process-level improvement, which extends detection to additional attack-associated nodes, the gain on resource entities comes from excluding benign nodes from the alert set. These complementary improvements account for much of JANUS's advantage in node-level detection.

Figure~\ref{fig:rq1-pr-curve} presents the precision--recall curves. JANUS achieves the highest F1 and MCC at the selected operating points, while THREATRACE obtains the highest AP.

\begin{table}[!htbp]
\centering
\begingroup
\papertablestyle{2pt}
\begin{tabularx}{\columnwidth}{@{}l*{3}{R}>{\hsize=1.2\hsize\linewidth=\hsize}R*{2}{>{\hsize=0.9\hsize\linewidth=\hsize}R}@{}}
\toprule
\rowcolor{TableHeaderBg}
\textbf{Method}
& \metrichead{Prec.}
& \metrichead{Rec.}
& \metrichead{F1}
& \metrichead{AUPRC}
& \metrichead{FPR}
& \metrichead{MCC} \\
\midrule
KubeLinter
& 0.0000 & 0.0000 & 0.0000
& 0.9772 & \textbf{0.0000} & 0.0000 \\
Kubescape
& \textbf{1.0000} & \textbf{0.4882} & \textbf{0.6561}
& \textbf{0.9893} & \textbf{0.0000} & \textbf{0.1393} \\
\bottomrule
\end{tabularx}
\endgroup
\caption{Comparison of static analysis tools.}
\label{tab:rq1-static}
\end{table}

Table~\ref{tab:rq1-static} presents the results for KubeLinter and Kubescape. Kubescape identifies 48.82\% of the labeled malicious processes without false-positive predictions, whereas KubeLinter produces no positive predictions at the selected threshold. KubeLinter relies on predefined configuration checks, which do not necessarily capture the authorization risks associated with malicious runtime activities. Consequently, attack-related processes may remain undetected when their configurations do not violate the enabled checks. The results highlight the limitations of configuration-based risk assessment for runtime intrusion detection. JANUS addresses this limitation by associating effective RBAC permissions with observed execution behavior.

\begin{observationbox}
\noindent\textbf{Answer to RQ1.}
JANUS achieves the highest F1 in binary permission--event matching and leads the node-level detection comparison in F1 and MCC. Its detection gains are most pronounced in improved recall for attack-associated processes and reduced false positives on resource nodes.
\end{observationbox}

\subsection{RQ2: Component Contributions}

\noindent\textbf{Setup.}
We evaluate the contributions of JANUS's key designs through ablations of the detection pipeline and the association objective. For detection, we consider three variants: (1) removing the permission branch, including static permission inputs and the association/fusion pathway; (2) replacing learned association with TF-IDF cosine similarity and uniform event-support weights; and (3) replacing gated residual fusion with concatenation followed by a linear projection. For association, we separately remove the multi-positive contrastive loss, binary cross-entropy loss, and ranking loss from the joint training objective.

All detection variants share the same runtime backbone, data split, and node-scoring rules, and are evaluated on 20,495 labeled test nodes using a single random seed. The detectors are trained exclusively on benign events, with thresholds set to the 85th percentile of their respective benign validation scores. For the association objective, all variants are independently retrained on the feature-disjoint data split using random seed 42. Matching thresholds are selected on the validation set and fixed for test evaluation.

\noindent\textbf{Results and Analysis.}
Table~\ref{tab:rq2-ablation} shows that permission conditioning primarily improves detection by suppressing false positives. Removing the permission branch reduces F1 from 0.3697 to 0.3404, while the number of detected malicious nodes changes only slightly (877 versus 873). In contrast, false positives increase from 2,332 to 2,722. The complete model therefore reduces false positives by 14.3\% relative to the variant without permissions, with nearly unchanged recall. This suggests that permission context helps distinguish benign activity from anomalous behavior rather than simply increasing detection sensitivity.

\begin{table}[!htbp]
\centering
\begingroup
\papertablestyle{1.2pt}
\begin{tabularx}{\columnwidth}{@{}L{0.36\columnwidth}*{3}{>{\hsize=0.95\hsize\linewidth=\hsize}R}>{\hsize=1.2\hsize\linewidth=\hsize}R>{\hsize=0.95\hsize\linewidth=\hsize}R@{}}
\toprule
\rowcolor{TableHeaderBg}
\textbf{Variant}
& \metrichead{Prec.}
& \metrichead{Rec.}
& \metrichead{F1}
& \metrichead{AUPRC}
& \metrichead{MCC} \\
\midrule
w/o permissions
& 0.2428 & 0.5687 & 0.3404
& 0.1822 & 0.2943 \\
w/ TF-IDF association
& 0.2323 & 0.5700 & 0.3301
& 0.1851 & 0.2837 \\
w/ concatenation fusion
& 0.2279 & 0.5668 & 0.3251
& 0.1791 & 0.2781 \\
\textbf{JANUS}
& \textbf{0.2733}
& \textbf{0.5713}
& \textbf{0.3697}
& \textbf{0.1963}
& \textbf{0.3247} \\
\bottomrule
\end{tabularx}
\endgroup
\caption{Node-level ablation results.}
\label{tab:rq2-ablation}
\end{table}

The effectiveness of permission conditioning also depends on how permission evidence is associated with runtime behavior and incorporated into provenance representations. Replacing learned association with TF-IDF or gated residual fusion with concatenation reduces F1 to 0.3301 and 0.3251, respectively, both below the variant without permissions. This indicates that permission information alone does not guarantee improved detection. Although the TF-IDF variant also adopts uniform event-support weights, applying uniform weights to the complete model changes MCC only marginally, from 0.3247 to 0.3241. The degradation therefore cannot be explained by uniform weighting alone.

\begin{center}
\begin{minipage}{\columnwidth}
\centering
\begingroup
\papertablestyle{3pt}
\begin{tabularx}{\columnwidth}{@{}L{0.44\columnwidth}*{3}{R}@{}}
\toprule
\rowcolor{TableHeaderBg}
\textbf{Variant}
& \metrichead{F1}
& \metrichead{MRR}
& \metrichead{AP} \\
\midrule
w/o contrastive loss
& 0.3158 & 0.2035 & 0.2375 \\
w/o BCE loss
& 0.9167 & \textbf{0.9250} & 0.9222 \\
w/o ranking loss
& 0.9000 & 0.9232 & 0.9171 \\
\textbf{JANUS}
& \textbf{0.9577} & 0.9234 & \textbf{0.9268} \\
\bottomrule
\end{tabularx}
\endgroup
\captionsetup{type=table,hypcap=false}
\caption{Ablation results for the association objective.}
\label{tab:rq2-loss-ablation}
\end{minipage}
\end{center}

We further examine the association objective in Table~\ref{tab:rq2-loss-ablation}. Removing the multi-positive contrastive loss causes the largest degradation, reducing F1 from 0.9577 to 0.3158 and AP from 0.9268 to 0.2375. This substantial decline, together with the reduction in MRR, highlights the importance of contrastive supervision in learning discriminative permission--event representations. In comparison, removing the binary cross-entropy or ranking loss reduces F1 to 0.9167 and 0.9000, respectively, while leaving MRR nearly unchanged. These results indicate that contrastive supervision is particularly important for cross-modal discrimination, whereas the other two objectives contribute to reliable binary matching.

\begin{observationbox}
\noindent\textbf{Answer to RQ2.}
Permission conditioning reduces node-level false positives by 14.3\% without substantially affecting recall. Learned association and gated fusion further improve detection, while contrastive supervision plays a central role in permission--event association.
\end{observationbox}

\subsection{RQ3: Hyperparameter Sensitivity}

\noindent\textbf{Setup.}
We investigate the sensitivity of JANUS to embedding dimension $d$, contrastive temperature $T$, hard-negative ratio $\rho$, inference neighbor count $N$, and the parameters of the association objective. We consider $d\in\{64,128,256\}$, $T\in\{0.07,0.12,0.20\}$, $\rho\in\{0.25,0.50,0.75\}$, and $N\in\{10,20,40\}$, with defaults of 128, 0.12, 0.50, and 20, respectively. The weights of the contrastive, binary cross-entropy, and ranking losses are varied over $\{0.25,0.5,1,2\}$, with a default of 1, while the ranking margin is varied over $\{0.05,0.15,0.30\}$, with a default of 0.15.

We vary one parameter at a time while keeping the others fixed. For $d$, $T$, and $\rho$, we retrain the association model and report validation-set matching performance using a threshold set to the 95th percentile of negative validation scores. For the loss weights and ranking margin, independently trained models are evaluated on the held-out association test set, with thresholds selected to maximize validation F1. The inference neighbor count $N$ is varied without retraining, and node-level detection thresholds are calibrated on benign validation data.

\begin{figure*}[!t]
\centering
\includegraphics[width=0.98\textwidth]{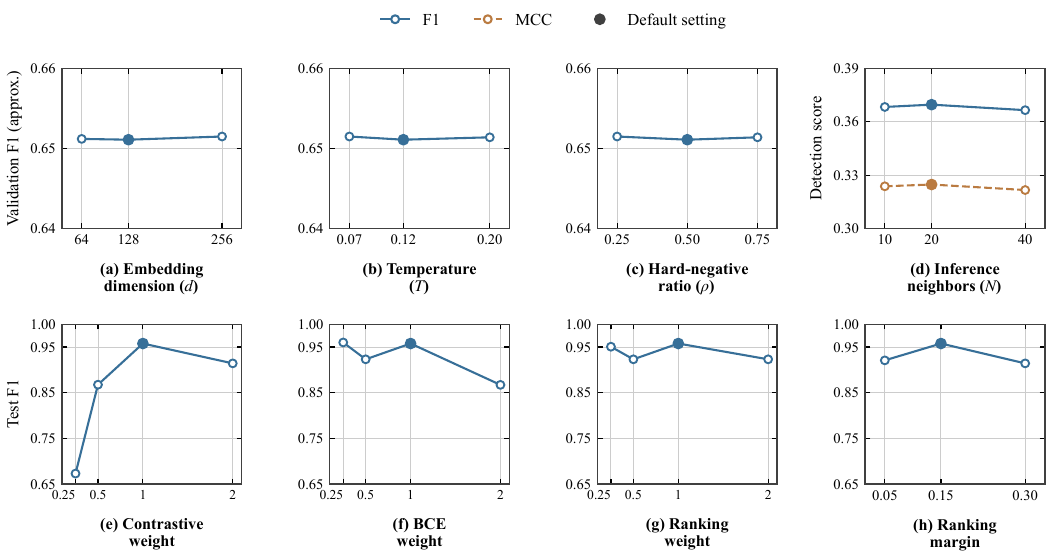}
\caption{Sensitivity of JANUS to association loss parameters and model settings. }
\label{fig:hyperparameter-sensitivity}
\end{figure*}

\noindent\textbf{Results and Analysis.}
Figure~\ref{fig:hyperparameter-sensitivity}(a)--(c) shows that association F1 ranges from 0.6511 to 0.6515 across the evaluated embedding dimensions, contrastive temperatures, and hard-negative ratios. These results indicate limited sensitivity to the three parameters within the evaluated ranges. Node-level detection is similarly insensitive to the inference neighbor count. As shown in Fig.~\ref{fig:hyperparameter-sensitivity}(d), varying $N$ from 10 to 40 yields F1 scores between 0.3666 and 0.3697 and MCC scores between 0.3216 and 0.3247, with both metrics reaching their highest values at $N=20$. Larger neighborhoods therefore provide no measurable improvement in this setting.

The association objective exhibits a different sensitivity pattern. Figure~\ref{fig:hyperparameter-sensitivity}(e)--(h) shows that reducing the contrastive loss weight from 1 to 0.25 decreases F1 from 0.9577 to 0.6728, whereas increasing it to 2 also reduces F1 to 0.9142. Consistent with the loss ablations in RQ2, these results highlight the importance of contrastive supervision and its relative contribution to the joint objective. The remaining parameters exhibit smaller but nonmonotonic effects. Increasing the BCE weight to 2 reduces F1 to 0.8671, while a weight of 0.25 yields 0.9600, slightly above the default configuration. Varying the ranking weight produces comparatively smaller changes, and the default ranking margin of 0.15 outperforms both 0.05 and 0.30. These results suggest that effective permission--event association depends on balancing the training objectives rather than maximizing the weight of any individual loss.

Increasing the embedding dimension or inference neighborhood beyond their default values yields no measurable performance improvement. For the association objective, the default loss weights and ranking margin achieve the highest F1 in three of the four parameter sweeps. Reducing the BCE weight from 1 to 0.25 produces a marginal F1 improvement of 0.0023. Overall, the selected configuration maintains strong performance without increasing model complexity or relying on larger loss weights.

\begin{observationbox}
\noindent\textbf{Answer to RQ3.}
JANUS exhibits limited sensitivity to the evaluated model and sampling parameters, while association performance depends more strongly on the contrastive loss weight. The default configuration achieves competitive performance without increasing representation size or inference neighborhood.
\end{observationbox}

\subsection{RQ4: Computational Efficiency}

\noindent\textbf{Setup.}
We evaluate the computational costs of permission--event association and runtime detection using 775,044 mapped benign validation events. Association is measured separately over three runs. For detection, we compare JANUS with a configuration that disables permission conditioning while retaining the same backbone and edge-feature dimensionality. Both configurations are evaluated in six paired runs using a single CPU thread pinned to a fixed core. We record cumulative processing time and cumulative peak process resident set size (RSS) at 20\%, 40\%, 60\%, 80\%, and 100\% of the event stream, reporting means and sample standard deviations across runs.

Static permission embeddings are reused during association, and detection uses precomputed association results. These measurements characterize individual processing stages rather than end-to-end online latency.

\begin{figure}[!htbp]
\centering
\includegraphics[width=\linewidth]{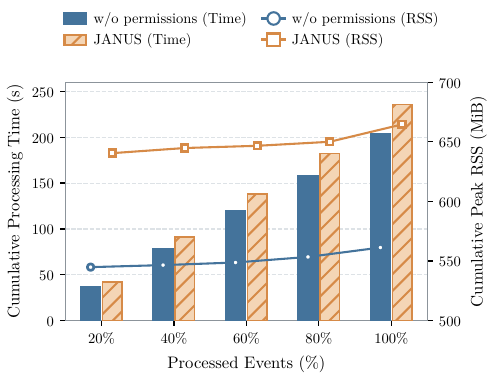}
\caption{Computational costs of JANUS with and without permission conditioning. Bars show cumulative processing time, and lines show cumulative peak RSS.}
\label{fig:rq4-cost}
\end{figure}

\noindent\textbf{Results and Analysis.}
The separately measured permission--event association stage takes 857.115~s to process 775,044 events. Figure~\ref{fig:rq4-cost} compares the detection costs of JANUS and the configuration without permission conditioning. At the final checkpoint, JANUS takes $235.689\pm1.704$~s, compared with $203.666\pm0.808$~s without permission conditioning. This corresponds to an additional 32.023~s (15.72\%), with a 95\% confidence interval of $[29.939,34.108]$~s for the paired difference.

As the number of processed events increases, both configurations exhibit increasing processing times. Although the absolute time difference grows from 5.58~s at 20\% to 32.02~s at 100\%, the relative overhead remains between 15.2\% and 15.7\% across all checkpoints. This suggests that permission conditioning introduces a roughly proportional processing cost over the evaluated workload, rather than an increasing relative penalty as more events are processed.

Memory consumption follows a different pattern. At the final checkpoint, mean peak RSS increases from $561.34\pm4.48$~MiB to $664.60\pm5.59$~MiB. However, the memory difference is already approximately 96~MiB at the first checkpoint and remains close to 100~MiB throughout the replay. Unlike the accumulating processing-time difference, the additional memory footprint changes only modestly with event volume. This suggests that much of the memory overhead is established early in detection rather than growing proportionally with the number of processed events.

Overall, the matched-backbone comparison shows that permission conditioning incurs a measurable but relatively consistent computational overhead across the evaluated event volumes. These measurements exclude association precomputation and therefore do not represent end-to-end online latency.

\begin{observationbox}
\noindent\textbf{Answer to RQ4.}
Permission conditioning increases mean detection time by 15.72\%, with a nearly constant relative overhead across event volumes. It also introduces an additional memory footprint that remains largely stable throughout the replay.
\end{observationbox}

\section{Discussion and Limitations}

\noindent\textbf{Quality of Training Data.}
JANUS learns normal workload behavior from benign execution traces, whose coverage may affect false-positive rates. For the main permission--event association evaluation, the training, validation, and test sets use distinct execution contexts and share neither raw event windows nor identical encoded event sequences. Nevertheless, all three sets cover the same 13 permission queries and are collected under controlled conditions. Generalization to unseen permissions, workloads, and Kubernetes environments therefore remains to be evaluated.

\noindent\textbf{Evidence of Permission Use.}
High-risk permissions may also be exercised by legitimate Kubernetes workloads, making permission use alone an unreliable indicator of compromise. In our ablation study, the complete permission-guided model reduces false positives by 14.3\% compared with the model trained without the permission branch, while maintaining comparable recall. However, unlike the controlled executions verified by Kubernetes Audit records during training, associations at detection time are inferred from system-level observations. Although process activities and client arguments provide useful clues, encrypted API requests may conceal the exact operations and resources involved. These associations therefore provide contextual evidence rather than definitive confirmation of permission use. Additional API-level observations could improve their reliability.

\noindent\textbf{Online Deployment.}
JANUS reuses precomputed permission embeddings and cached association evidence during detection. In the single-core evaluation, the detection stage processes 775,044 benign events in $235.689\pm1.704$~s. Since permission--event association is measured separately, these results do not establish the end-to-end latency of an integrated online system. Moreover, changes to RBAC roles or bindings may invalidate cached permission evidence, requiring timely updates. Supporting incremental authorization updates and evaluating continuous detection under dynamic authorization states remain future work.

\section{Related Work}

\noindent\textbf{Kubernetes Permission Security Analysis.}
Container security guidance identifies excessive privileges and insecure configurations as deployment risks~\citep{nist_container}. Existing studies investigate excessive permissions and exploitable authorization paths in Kubernetes. Yang et al. demonstrate how excessive permissions granted to third-party applications can be exploited to compromise an entire cluster~\citep{excessive_permissions}. EPScan combines configuration analysis with pod-oriented program analysis to identify exploitable excessive RBAC permissions~\citep{epscan}, while KubiScan examines roles and bindings to identify risky grants~\citep{kubiscan}.

KubeSec examines takeover risks introduced by third-party Kubernetes applications through permission configurations and component code dependencies~\citep{kubesec}. Yang et al. analyze implicit permissions arising from interactions among Kubernetes resources, combining static analysis with language-model-assisted reasoning to identify associated risks~\citep{implicit_permissions}. KubeKeeper protects Secrets using encryption and admission-controlled access to decrypted values~\citep{kubekeeper}. These studies address authorization risks and secret protection, whereas JANUS relates effective RBAC permissions to observed execution as intrusion-detection context.

\noindent\textbf{Provenance-based Intrusion Detection.}
Provenance-based approaches analyze dependencies among system entities to identify suspicious activities~\citep{provenance_challenges}. UNICORN models long-term execution behavior using graph sketches~\citep{unicorn}, while THREATRACE learns benign entity roles to detect node-level anomalies~\citep{threatrace}. MAGIC combines masked graph representation learning with outlier detection~\citep{magic}, whereas FLASH integrates semantic and structural information with embedding reuse to improve detection efficiency~\citep{flash}. KAIROS models temporal dependencies for intrusion detection and attack investigation~\citep{kairos}, while ORTHRUS focuses on node-level detection and improving attribution quality~\citep{orthrus}.

Other studies explore complementary provenance analysis techniques. ProvG-Searcher searches for known attack behaviors in provenance graphs~\citep{provg_searcher}, while RDD investigates real-time detection through graph matching~\citep{rdd}. CAPTAIN introduces adaptive rule-based detection through differentiable tag propagation~\citep{captain}.

Online systems address the cost of provenance analysis in different ways. NODLINK uses an online approximation to the Steiner tree problem to recover compact attack-related subgraphs~\citep{nodlink}. TAPAS combines recurrent process representations with task-guided graph segmentation for efficient online detection~\citep{tapas}. KnowHow maps cyber threat intelligence knowledge to system events through general indicators of compromise and temporal reasoning~\citep{knowhow}.

Recent work also examines supervision and evaluation. AutoLabel builds fine-grained attack labels by combining provenance graphs with auxiliary execution information~\citep{autolabel}. DUPIN pretrains on provenance data and then uses a few labeled attack examples for adaptation~\citep{dupin}. Bilot et al. evaluate eight provenance-based detectors in a unified framework and identify limitations affecting practical detection and evaluation~\citep{pids_evaluation}. These studies provide complementary perspectives on labeling, attack supervision, and comparable evaluation. JANUS focuses on associating effective Kubernetes RBAC permissions with execution evidence and using that evidence to condition node-level detection.

\noindent\textbf{Cross-Modal Representation Learning.}
Contrastive learning learns representations by contrasting matching and mismatching views. SimCLR studies augmentation-based visual contrastive learning~\citep{simclr}, while GraphCL learns representations from augmented graphs~\citep{graphcl}. Supervised contrastive learning allows examples sharing a class to serve as multiple positives~\citep{supcon}. These objectives provide context for multi-positive learning; JANUS groups positives by verified permission semantics.

CLIP aligns separate image and text encoders through contrastive supervision~\citep{clip}, and FILIP uses token-level late interaction for fine-grained image--text matching~\citep{filip}. ConGraT jointly learns graph-node and text embeddings~\citep{congrat}. CAME combines cross-modal contrastive alignment with modality-aware expert fusion for multimodal graphs~\citep{came}. Gated fusion is also studied in the Multimodal Adaptation Gate, which conditions pretrained language representations on visual and acoustic signals~\citep{mag}. These studies concern image--text, graph--text, or multimodal language tasks. JANUS applies cross-modal association to verified permission--event pairs and enforces temporal constraints when fusing permission evidence into provenance representations.

\section{Conclusion}

We presented JANUS, a permission-guided intrusion detection approach that connects Kubernetes authorization context with system-level provenance. JANUS recovers effective RBAC permissions, learns their associations with runtime events using Audit-verified supervision, and incorporates historical permission evidence into provenance representations for anomaly detection. By combining permission-use evidence with behavioral anomaly detection, JANUS provides authorization context for identifying suspicious workload activities.

On the node-level test set, JANUS achieves an F1 score of 0.3697 and an MCC of 0.3247. Compared with the model trained without the permission branch, JANUS reduces false positives by 14.3\% while maintaining comparable recall. These results demonstrate the potential of permission-guided detection to reduce false alarms in Kubernetes environments.

\begingroup
% Compact spacing for the end-matter headings.
\makeatletter
\renewcommand\section{\@startsection{section}{1}{\z@}%
  {6\p@ \@plus 2\p@ \@minus 1\p@}%
  {3\p@ \@plus 1\p@ \@minus 1\p@}%
  {\normalsize\bfseries\boldmath}}
\makeatother
\section*{Acknowledgements}

This work is supported in part by the Hainan Province Science and Technology Special Fund under Grant ZDYF2024GXJS008; in part by the National Natural Science Foundation of China under Grants U22B2027, U2468204, U25A20424, 62503362, and 62672346; in part by the Joint Special Project of the Beijing-Tianjin-Hebei Natural Science Foundation under Grant 25JJJJC0033.

\section*{Declaration of generative AI and AI-assisted technologies in the writing process}

During the preparation of this manuscript, the authors used ChatGPT (OpenAI) to assist with manuscript organization, language refinement, and editorial review of technical descriptions and experimental discussions. The authors reviewed and revised the AI-assisted content and take full responsibility for the final content of the manuscript.
\par\endgroup

\bibliography{references}
\end{document}